\documentclass[conference,10pt,twoside,twocolumn]{IEEEtran}

\usepackage{cite}
\usepackage[T1]{fontenc}
\usepackage{graphicx}
\usepackage{amssymb}
\usepackage{amsmath}
\usepackage{amsthm}
\usepackage{subfigure}

\usepackage{booktabs}
\usepackage{multirow}
\usepackage{microtype}
\usepackage{balance}
\usepackage{xcolor}
\usepackage{balance}

\usepackage{tikz}
\usepackage{pgfplots}
\pgfplotsset{compat=newest}
\pgfplotsset{plot coordinates/math parser=false}
\newlength\figureheight
\newlength\figurewidth

\usepackage{amssymb}
\usepackage{amsfonts}
\usepackage{mathrsfs}
\usepackage{xspace}
\usepackage{bm}
\usepackage{upgreek}

\newcommand{\safemath}[2]{\newcommand{#1}{\ensuremath{#2}\xspace}}

\safemath{\bma}{\mathbf{a}}
\safemath{\bmb}{\mathbf{b}}
\safemath{\bmc}{\mathbf{c}}
\safemath{\bmd}{\mathbf{d}}
\safemath{\bme}{\mathbf{e}}
\safemath{\bmf}{\mathbf{f}}
\safemath{\bmg}{\mathbf{g}}
\safemath{\bmh}{\mathbf{h}}
\safemath{\bmi}{\mathbf{i}}
\safemath{\bmj}{\mathbf{j}}
\safemath{\bmk}{\mathbf{k}}
\safemath{\bml}{\mathbf{l}}
\safemath{\bmm}{\mathbf{m}}
\safemath{\bmn}{\mathbf{n}}
\safemath{\bmo}{\mathbf{o}}
\safemath{\bmp}{\mathbf{p}}
\safemath{\bmq}{\mathbf{q}}
\safemath{\bmr}{\mathbf{r}}
\safemath{\bms}{\mathbf{s}}
\safemath{\bmt}{\mathbf{t}}
\safemath{\bmu}{\mathbf{u}}
\safemath{\bmv}{\mathbf{v}}
\safemath{\bmw}{\mathbf{w}}
\safemath{\bmx}{\mathbf{x}}
\safemath{\bmy}{\mathbf{y}}
\safemath{\bmz}{\mathbf{z}}
\safemath{\bmzero}{\mathbf{0}}
\safemath{\bmone}{\mathbf{1}}

\bmdefine{\biad}{a}
\bmdefine{\bibd}{b}
\bmdefine{\bicd}{c}
\bmdefine{\bidd}{d}
\bmdefine{\bied}{e}
\bmdefine{\bifd}{f}
\bmdefine{\bigd}{g}
\bmdefine{\bihd}{h}
\bmdefine{\biid}{i}
\bmdefine{\bijd}{j}
\bmdefine{\bikd}{k}
\bmdefine{\bild}{l}
\bmdefine{\bimd}{m}
\bmdefine{\bind}{n}
\bmdefine{\biod}{o}
\bmdefine{\bipd}{p}
\bmdefine{\biqd}{q}
\bmdefine{\bird}{r}
\bmdefine{\bisd}{s}
\bmdefine{\bitd}{t}
\bmdefine{\biud}{u}
\bmdefine{\bivd}{v}
\bmdefine{\biwd}{w}
\bmdefine{\bixd}{x}
\bmdefine{\biyd}{y}
\bmdefine{\bizd}{z}

\bmdefine{\bixid}{\xi}
\bmdefine{\bilambdad}{\lambda}
\bmdefine{\bimud}{\mu}
\bmdefine{\bithetad}{\theta}
\bmdefine{\biphid}{\phi}
\bmdefine{\bideltad}{\delta}

\safemath{\bmia}{\biad}
\safemath{\bmib}{\bibd}
\safemath{\bmic}{\bicd}
\safemath{\bmid}{\bidd}
\safemath{\bmie}{\bied}
\safemath{\bmif}{\bifd}
\safemath{\bmig}{\bigd}
\safemath{\bmih}{\bihd}
\safemath{\bmii}{\biid}
\safemath{\bmij}{\bijd}
\safemath{\bmik}{\bikd}
\safemath{\bmil}{\bild}
\safemath{\bmim}{\bimd}
\safemath{\bmin}{\bind}
\safemath{\bmio}{\biod}
\safemath{\bmip}{\bipd}
\safemath{\bmiq}{\biqd}
\safemath{\bmir}{\bird}
\safemath{\bmis}{\bisd}
\safemath{\bmit}{\bitd}
\safemath{\bmiu}{\biud}
\safemath{\bmiv}{\bivd}
\safemath{\bmiw}{\biwd}
\safemath{\bmix}{\bixd}
\safemath{\bmiy}{\biyd}
\safemath{\bmiz}{\bizd}

\safemath{\bmxi}{\bixid}
\safemath{\bmlambda}{\bilambdad}
\safemath{\bmmu}{\bimud}
\safemath{\bmtheta}{\bithetad}
\safemath{\bmphi}{\biphid}
\safemath{\bmdelta}{\bideltad}

\safemath{\bA}{\mathbf{A}}
\safemath{\bB}{\mathbf{B}}
\safemath{\bC}{\mathbf{C}}
\safemath{\bD}{\mathbf{D}}
\safemath{\bE}{\mathbf{E}}
\safemath{\bF}{\mathbf{F}}
\safemath{\bG}{\mathbf{G}}
\safemath{\bH}{\mathbf{H}}
\safemath{\bI}{\mathbf{I}}
\safemath{\bJ}{\mathbf{J}}
\safemath{\bK}{\mathbf{K}}
\safemath{\bL}{\mathbf{L}}
\safemath{\bM}{\mathbf{M}}
\safemath{\bN}{\mathbf{N}}
\safemath{\bO}{\mathbf{O}}
\safemath{\bP}{\mathbf{P}}
\safemath{\bQ}{\mathbf{Q}}
\safemath{\bR}{\mathbf{R}}
\safemath{\bS}{\mathbf{S}}
\safemath{\bT}{\mathbf{T}}
\safemath{\bU}{\mathbf{U}}
\safemath{\bV}{\mathbf{V}}
\safemath{\bW}{\mathbf{W}}
\safemath{\bX}{\mathbf{X}}
\safemath{\bY}{\mathbf{Y}}
\safemath{\bZ}{\mathbf{Z}}

\safemath{\bZero}{\mathbf{0}}
\safemath{\bOne}{\mathbf{1}}
\safemath{\bDelta}{\mathbf{\Delta}}
\safemath{\bLambda}{\mathbf{\UpLambda}}
\safemath{\bPhi}{\mathbf{\Upphi}}
\safemath{\bSigma}{\mathbf{\Upsigma}}
\safemath{\bOmega}{\mathbf{\Upomega}}
\safemath{\bTheta}{\mathbf{\Uptheta}}

\bmdefine{\biAd}{A}
\bmdefine{\biBd}{B}
\bmdefine{\biCd}{C}
\bmdefine{\biDd}{D}
\bmdefine{\biEd}{E}
\bmdefine{\biFd}{F}
\bmdefine{\biGd}{G}
\bmdefine{\biHd}{H}
\bmdefine{\biId}{I}
\bmdefine{\biJd}{J}
\bmdefine{\biKd}{K}
\bmdefine{\biLd}{L}
\bmdefine{\biMd}{M}
\bmdefine{\biOd}{N}
\bmdefine{\biPd}{O}
\bmdefine{\biQd}{P}
\bmdefine{\biRd}{R}
\bmdefine{\biSd}{S}
\bmdefine{\biTd}{T}
\bmdefine{\biUd}{U}
\bmdefine{\biVd}{V}
\bmdefine{\biWd}{W}
\bmdefine{\biXd}{X}
\bmdefine{\biYd}{Y}
\bmdefine{\biZd}{Z}

\bmdefine{\biDelta}{\Delta}
\bmdefine{\biLambda}{\Lambda}
\bmdefine{\biPhi}{\Phi}
\bmdefine{\biSigma}{\Sigma}
\bmdefine{\biOmega}{\Omega}
\bmdefine{\biTheta}{\Theta}

\safemath{\bimA}{\biAd}
\safemath{\bimB}{\biBd}
\safemath{\bimC}{\biCd}
\safemath{\bimD}{\biDd}
\safemath{\bimE}{\biEd}
\safemath{\bimF}{\biFd}
\safemath{\bimG}{\biGd}
\safemath{\bimH}{\biHd}
\safemath{\bimI}{\biId}
\safemath{\bimJ}{\biJd}
\safemath{\bimK}{\biKd}
\safemath{\bimL}{\biLd}
\safemath{\bimM}{\biMd}
\safemath{\bimN}{\biNd}
\safemath{\bimO}{\biOd}
\safemath{\bimP}{\biPd}
\safemath{\bimQ}{\biQd}
\safemath{\bimR}{\biRd}
\safemath{\bimS}{\biSd}
\safemath{\bimT}{\biTd}
\safemath{\bimU}{\biUd}
\safemath{\bimV}{\biVd}
\safemath{\bimW}{\biWd}
\safemath{\bimX}{\biXd}
\safemath{\bimY}{\biYd}
\safemath{\bimZ}{\biZd}

\safemath{\bimDelta}{\biDelta}
\safemath{\bimLambda}{\biLambda}
\safemath{\bimPhi}{\biPhi}
\safemath{\bimSigma}{\biSigma}
\safemath{\bimOmega}{\biOmega}
\safemath{\bimTheta}{\biTheta}

\safemath{\setA}{\mathcal{A}}
\safemath{\setB}{\mathcal{B}}
\safemath{\setC}{\mathcal{C}}
\safemath{\setD}{\mathcal{D}}
\safemath{\setE}{\mathcal{E}}
\safemath{\setF}{\mathcal{F}}
\safemath{\setG}{\mathcal{G}}
\safemath{\setH}{\mathcal{H}}
\safemath{\setI}{\mathcal{I}}
\safemath{\setJ}{\mathcal{J}}
\safemath{\setK}{\mathcal{K}}
\safemath{\setL}{\mathcal{L}}
\safemath{\setM}{\mathcal{M}}
\safemath{\setN}{\mathcal{N}}
\safemath{\setO}{\mathcal{O}}
\safemath{\setP}{\mathcal{P}}
\safemath{\setQ}{\mathcal{Q}}
\safemath{\setR}{\mathcal{R}}
\safemath{\setS}{\mathcal{S}}
\safemath{\setT}{\mathcal{T}}
\safemath{\setU}{\mathcal{U}}
\safemath{\setV}{\mathcal{V}}
\safemath{\setW}{\mathcal{W}}
\safemath{\setX}{\mathcal{X}}
\safemath{\setY}{\mathcal{Y}}
\safemath{\setZ}{\mathcal{Z}}
\safemath{\emptySet}{\varnothing}

\safemath{\colA}{\mathscr{A}}
\safemath{\colB}{\mathscr{B}}
\safemath{\colC}{\mathscr{C}}
\safemath{\colD}{\mathscr{D}}
\safemath{\colE}{\mathscr{E}}
\safemath{\colF}{\mathscr{F}}
\safemath{\colG}{\mathscr{G}}
\safemath{\colH}{\mathscr{H}}
\safemath{\colI}{\mathscr{I}}
\safemath{\colJ}{\mathscr{J}}
\safemath{\colK}{\mathscr{K}}
\safemath{\colL}{\mathscr{L}}
\safemath{\colM}{\mathscr{M}}
\safemath{\colN}{\mathscr{N}}
\safemath{\colO}{\mathscr{O}}
\safemath{\colP}{\mathscr{P}}
\safemath{\colQ}{\mathscr{Q}}
\safemath{\colR}{\mathscr{R}}
\safemath{\colS}{\mathscr{S}}
\safemath{\colT}{\mathscr{T}}
\safemath{\colU}{\mathscr{U}}
\safemath{\colV}{\mathscr{V}}
\safemath{\colW}{\mathscr{W}}
\safemath{\colX}{\mathscr{X}}
\safemath{\colY}{\mathscr{Y}}
\safemath{\colZ}{\mathscr{Z}}

\safemath{\opA}{\mathbb{A}}
\safemath{\opB}{\mathbb{B}}
\safemath{\opC}{\mathbb{C}}
\safemath{\opD}{\mathbb{D}}
\safemath{\opE}{\mathbb{E}}
\safemath{\opF}{\mathbb{F}}
\safemath{\opG}{\mathbb{G}}
\safemath{\opH}{\mathbb{H}}
\safemath{\opI}{\mathbb{I}}
\safemath{\opJ}{\mathbb{J}}
\safemath{\opK}{\mathbb{K}}
\safemath{\opL}{\mathbb{L}}
\safemath{\opM}{\mathbb{M}}
\safemath{\opN}{\mathbb{N}}
\safemath{\opO}{\mathbb{O}}
\safemath{\opP}{\mathbb{P}}
\safemath{\opQ}{\mathbb{Q}}
\safemath{\opR}{\mathbb{R}}
\safemath{\opS}{\mathbb{S}}
\safemath{\opT}{\mathbb{T}}
\safemath{\opU}{\mathbb{U}}
\safemath{\opV}{\mathbb{V}}
\safemath{\opW}{\mathbb{W}}
\safemath{\opX}{\mathbb{X}}
\safemath{\opY}{\mathbb{Y}}
\safemath{\opZ}{\mathbb{Z}}
\safemath{\opZero}{\mathbb{O}}
\safemath{\identityop}{\opI}

\safemath{\veca}{\bma}
\safemath{\vecb}{\bmb}
\safemath{\vecc}{\bmc}
\safemath{\vecd}{\bmd}
\safemath{\vece}{\bme}
\safemath{\vecf}{\bmf}
\safemath{\vecg}{\bmg}
\safemath{\vech}{\bmh}
\safemath{\veci}{\bmi}
\safemath{\vecj}{\bmj}
\safemath{\veck}{\bmk}
\safemath{\vecl}{\bml}
\safemath{\vecm}{\bmm}
\safemath{\vecn}{\bmn}
\safemath{\veco}{\bmo}
\safemath{\vecp}{\bmp}
\safemath{\vecq}{\bmq}
\safemath{\vecr}{\bmr}
\safemath{\vecs}{\bms}
\safemath{\vect}{\bmt}
\safemath{\vecu}{\bmu}
\safemath{\vecv}{\bmv}
\safemath{\vecw}{\bmw}
\safemath{\vecx}{\bmx}
\safemath{\vecy}{\bmy}
\safemath{\vecz}{\bmz}

\safemath{\veczero}{\bmzero}
\safemath{\vecone}{\bmone}
\safemath{\vecxi}{\bmxi}
\safemath{\veclambda}{\bmlambda}
\safemath{\vecmu}{\bmmu}
\safemath{\vectheta}{\bmtheta}
\safemath{\vecphi}{\bmphi}
\safemath{\vecdelta}{\bmdelta}

\safemath{\matA}{\bA}
\safemath{\matB}{\bB}
\safemath{\matC}{\bC}
\safemath{\matD}{\bD}
\safemath{\matE}{\bE}
\safemath{\matF}{\bF}
\safemath{\matG}{\bG}
\safemath{\matH}{\bH}
\safemath{\matI}{\bI}
\safemath{\matJ}{\bJ}
\safemath{\matK}{\bK}
\safemath{\matL}{\bL}
\safemath{\matM}{\bM}
\safemath{\matN}{\bN}
\safemath{\matO}{\bO}
\safemath{\matP}{\bP}
\safemath{\matQ}{\bQ}
\safemath{\matR}{\bR}
\safemath{\matS}{\bS}
\safemath{\matT}{\bT}
\safemath{\matU}{\bU}
\safemath{\matV}{\bV}
\safemath{\matW}{\bW}
\safemath{\matX}{\bX}
\safemath{\matY}{\bY}
\safemath{\matZ}{\bZ}
\safemath{\matzero}{\bmzero}

\safemath{\matDelta}{\bDelta}
\safemath{\matLambda}{\bLambda}
\safemath{\matPhi}{\bPhi}
\safemath{\matSigma}{\bSigma}
\safemath{\matOmega}{\bOmega}
\safemath{\matTheta}{\bTheta}

\safemath{\matidentity}{\matI}
\safemath{\matone}{\matO}

\safemath{\rnda}{A}
\safemath{\rndb}{B}
\safemath{\rndc}{C}
\safemath{\rndd}{D}
\safemath{\rnde}{E}
\safemath{\rndf}{F}
\safemath{\rndg}{G}
\safemath{\rndh}{H}
\safemath{\rndi}{I}
\safemath{\rndj}{J}
\safemath{\rndk}{K}
\safemath{\rndl}{L}
\safemath{\rndm}{M}
\safemath{\rndn}{N}
\safemath{\rndo}{O}
\safemath{\rndp}{P}
\safemath{\rndq}{Q}
\safemath{\rndr}{R}
\safemath{\rnds}{S}
\safemath{\rndt}{T}
\safemath{\rndu}{U}
\safemath{\rndv}{V}
\safemath{\rndw}{W}
\safemath{\rndx}{X}
\safemath{\rndy}{Y}
\safemath{\rndz}{Z}

\safemath{\rveca}{\bimA}
\safemath{\rvecb}{\bimB}
\safemath{\rvecc}{\bimC}
\safemath{\rvecd}{\bimD}
\safemath{\rvece}{\bimE}
\safemath{\rvecf}{\bimF}
\safemath{\rvecg}{\bimG}
\safemath{\rvech}{\bimH}
\safemath{\rveci}{\bimI}
\safemath{\rvecj}{\bimJ}
\safemath{\rveck}{\bimK}
\safemath{\rvecl}{\bimL}
\safemath{\rvecm}{\bimM}
\safemath{\rvecn}{\bimN}
\safemath{\rveco}{\bomO}
\safemath{\rvecp}{\bimP}
\safemath{\rvecq}{\bimQ}
\safemath{\rvecr}{\bimR}
\safemath{\rvecs}{\bimS}
\safemath{\rvect}{\bimT}
\safemath{\rvecu}{\bimU}
\safemath{\rvecv}{\bimV}
\safemath{\rvecw}{\bimW}
\safemath{\rvecx}{\bimX}
\safemath{\rvecy}{\bimY}
\safemath{\rvecz}{\bimZ}

\safemath{\rvecxi}{\bmxi}
\safemath{\rveclambda}{\bmlambda}
\safemath{\rvecmu}{\bmmu}
\safemath{\rvectheta}{\bmtheta}
\safemath{\rvecphi}{\bmphi}

\safemath{\rmatA}{\bimA}
\safemath{\rmatB}{\bimB}
\safemath{\rmatC}{\bimC}
\safemath{\rmatD}{\bimD}
\safemath{\rmatE}{\bimE}
\safemath{\rmatF}{\bimF}
\safemath{\rmatG}{\bimG}
\safemath{\rmatH}{\bimH}
\safemath{\rmatI}{\bimI}
\safemath{\rmatJ}{\bimJ}
\safemath{\rmatK}{\bimK}
\safemath{\rmatL}{\bimL}
\safemath{\rmatM}{\bimM}
\safemath{\rmatN}{\bimN}
\safemath{\rmatO}{\bimO}
\safemath{\rmatP}{\bimP}
\safemath{\rmatQ}{\bimQ}
\safemath{\rmatR}{\bimR}
\safemath{\rmatS}{\bimS}
\safemath{\rmatT}{\bimT}
\safemath{\rmatU}{\bimU}
\safemath{\rmatV}{\bimV}
\safemath{\rmatW}{\bimW}
\safemath{\rmatX}{\bimX}
\safemath{\rmatY}{\bimY}
\safemath{\rmatZ}{\bimZ}

\safemath{\rmatDelta}{\bimDelta}
\safemath{\rmatLambda}{\bimLambda}
\safemath{\rmatPhi}{\bimPhi}
\safemath{\rmatSigma}{\bimSigma}
\safemath{\rmatOmega}{\bimOmega}
\safemath{\rmatTheta}{\bimTheta}

\usepackage{amssymb}
\usepackage{amsfonts}
\usepackage{mathrsfs}
\usepackage{xspace}
\usepackage{bm}
\usepackage{fancyref}
\usepackage{textcomp}

\usepackage{multirow}
\usepackage{stmaryrd}

\newenvironment{textbmatrix}{	\setlength{\arraycolsep}{2.5pt}%
								\big[\begin{matrix}}{\end{matrix}\big]%
								\raisebox{0.08ex}{\vphantom{M}}}

\def\be{\begin{equation}}
\def\ee{\end{equation}}
\def\een{\nonumber \end{equation}}
\def\mat{\begin{bmatrix}}
\def\emat{\end{bmatrix}}
\def\btm{\begin{textbmatrix}}
\def\etm{\end{textbmatrix}}

\def\ba#1\ea{\begin{align}#1\end{align}}
\def\bas#1\eas{\begin{align*}#1\end{align*}}
\def\bs#1\es{\begin{split}#1\end{split}}
\def\bg#1\eg{\begin{gather}#1\end{gather}}
\def\bml#1\eml{\begin{multline}#1\end{multline}}
\def\bi#1\ei{\begin{itemize}#1\end{itemize}}

\safemath{\dirac}{\delta}					% Dirac delta
\safemath{\krond}{\dirac}					% Kronecker delta
\safemath{\upto}{\uparrow}
\safemath{\downto}{\downarrow}
\safemath{\iu}{j}							% imaginary unit
\safemath{\ev}{\lambda}						% eigenvalue
\safemath{\hilseqspace}{l^{2}}				% Hilbert sequence space
\newcommand{\banachfunspace}[1]{\setL^{#1}}	% Banach function space
\safemath{\hilfunspace}{\banachfunspace{2}}	% Hilbert function space
\safemath{\SNR}{\textit{SNR}} 				% signal to noise ratio
\safemath{\PAR}{\textit{PAR}} 				% signal to noise ratio
\safemath{\No}{N_0}							% noise spectral density
\safemath{\Es}{E_s}							% energy per symbol
\safemath{\Eb}{E_b}							% energy per bit
\safemath{\EbNo}{\frac{\Eb}{\No}}
\safemath{\EsNo}{\frac{\Es}{\No}}

\DeclareMathOperator{\CHop}{\ensuremath{\opH}} % channel operator
\safemath{\tvir}{\rndh_{\CHop}}				% time-varying impulse response
\safemath{\tvtf}{\rndl_{\CHop}}				% 	-''- transfer function
\safemath{\spf}{\rnds_{\CHop}}				% spreading function
\safemath{\bff}{H_{\CHop}}					% bi-freuqency function

\safemath{\ircf}{r_{h}}						% impulse response correlation fn.
\safemath{\tftvcf}{r_{s}}					% scattering function
\safemath{\tfcf}{r_{l}}						% time-frequency correlation fn.
\safemath{\bfcf}{r_{H}}						% bi-frequency correlation fn.

\safemath{\tcorr}{c_h}						% time-correlation function
\safemath{\scf}{c_{s}}						% spreading function
\safemath{\tfcorr}{c_{l}}					% transfer-function correlation
\safemath{\fcorr}{c_{H}}						% frequency-correlation function

\safemath{\mi}{I}							% mutual information
\safemath{\capacity}{C}						% capacity

\safemath{\normal}{\mathcal{N}}			% normal distribution
\safemath{\jpg}{\mathcal{CN}}			% jointly proper Gaussian
\safemath{\mchain}{\leftrightarrow}		% Markov chain
\safemath{\dB}{\,\mathrm{dB}}
\safemath{\dBm}{\,\mathrm{dBm}}
\safemath{\Hz}{\,\mathrm{Hz}}
\safemath{\kHz}{\,\mathrm{kHz}}
\safemath{\MHz}{\,\mathrm{MHz}}
\safemath{\GHz}{\,\mathrm{GHz}}
\safemath{\s}{\,\mathrm{s}}
\safemath{\ms}{\,\mathrm{ms}}
\safemath{\mus}{\,\mathrm{\text{\textmu}s}}
\safemath{\ns}{\,\mathrm{ns}}
\safemath{\ps}{\,\mathrm{ps}}
\safemath{\meter}{\,\mathrm{m}}
\safemath{\mm}{\,\mathrm{mm}}
\safemath{\cm}{\,\mathrm{cm}}
\safemath{\m}{\,\mathrm{m}}
\safemath{\W}{\,\mathrm{W}}
\safemath{\mW}{\, \mathrm{mW}}
\safemath{\J}{\,\mathrm{J}}
\safemath{\K}{\,\mathrm{K}}
\safemath{\bit}{\,\mathrm{bit}}
\safemath{\nat}{\,\mathrm{nat}}

\safemath{\define}{\triangleq}			% definition

\safemath{\equivalent}{\sim}
\safemath{\distas}{\sim}					% distributed according to
\safemath{\sdiff}{\Delta}				% symmetric set difference

\safemath{\reals}{\mathbb{R}}
\safemath{\positivereals}{\reals_{+}}
\safemath{\integers}{\mathbb{Z}}
\safemath{\posint}{\integers_{+}}
\safemath{\naturals}{\mathbb{N}}
\safemath{\posnaturals}{\naturals_{+}}
\safemath{\complexset}{\mathbb{C}}
\safemath{\rationals}{\mathbb{Q}}

\newcommand*{\fancyrefapplabelprefix}{app}		% Appendix
\newcommand*{\fancyrefthmlabelprefix}{thm}		% Theorem
\newcommand*{\fancyreflemlabelprefix}{lem}		% Lemma
\newcommand*{\fancyrefcorlabelprefix}{cor}		% Corollary
\newcommand*{\fancyrefdeflabelprefix}{def}		% Definition
\newcommand*{\fancyrefproplabelprefix}{prop}		% Proposition
\newcommand*{\fancyrefexmpllabelprefix}{exmpl}
\newcommand*{\fancyrefalglabelprefix}{alg}		% Algorithm
\newcommand*{\fancyreftbllabelprefix}{tbl}		% Algorithm

\frefformat{vario}{\fancyrefseclabelprefix}{Sec.~#1}
\frefformat{vario}{\fancyrefthmlabelprefix}{Thm.~#1}
\frefformat{vario}{\fancyreftbllabelprefix}{Tbl.~#1}
\frefformat{vario}{\fancyreflemlabelprefix}{Lem.~#1}
\frefformat{vario}{\fancyrefcorlabelprefix}{Cor.~#1}
\frefformat{vario}{\fancyrefdeflabelprefix}{Def.~#1}
\frefformat{vario}{\fancyreffiglabelprefix}{Fig.~#1}
\frefformat{vario}{\fancyrefapplabelprefix}{App.~#1}
\frefformat{vario}{\fancyrefeqlabelprefix}{(#1)}
\frefformat{vario}{\fancyrefproplabelprefix}{Prop.~#1}
\frefformat{vario}{\fancyrefexmpllabelprefix}{Ex.~#1}
\frefformat{vario}{\fancyrefalglabelprefix}{Alg.~#1}

\safemath{\dictab}{[\,\dicta\,\,\dictb\,]}

\safemath{\ysig}{\bmy}
\safemath{\ysighat}{\hat{\ysig}}
\safemath{\ysigdim}{M}
\safemath{\xsig}{\bmx}
\safemath{\xsigdim}{N}
\safemath{\nx}{n_x}
\safemath{\zsig}{\bmz}
\safemath{\zsigdim}{\ysigdim}
\safemath{\rsig}{\bmr}
\safemath{\Adict}{\bA}
\safemath{\Adicttilde}{\widetilde{\Adict}}
\safemath{\Adictdim}{\outputdim\times\xsigdim}
\safemath{\avec}{\bma}
\safemath{\avectilde}{\tilde{\avec}}
\safemath{\Bdict}{\bB}
\safemath{\Bdicttilde}{\widetilde{\Bdict}}
\safemath{\Cdict}{\bC}
\safemath{\cvec}{\bmc}
\safemath{\Ddict}{\bD}
\safemath{\Ddictdim}{\ysigdim\times\xsigdim}
\safemath{\dvec}{\bmd}
\safemath{\Ddicttilde}{\widetilde{\bD}}
\safemath{\Bonb}{\bB}
\safemath{\bvec}{\bmb}
\safemath{\Bonbdim}{\ysigdim\times\ysigdim}
\safemath{\noise}{\bmn}
\safemath{\noisedim}{\ysigim}
\safemath{\err}{\bme}
\safemath{\errdim}{\ysigdim}
\safemath{\errset}{\setE}
\safemath{\nerr}{n_e}
\safemath{\delop}{\bP_\errset}
\safemath{\delopc}{\bP_{{\errset}^c}}

\safemath{\cplxi}{\imath}
\safemath{\cplxj}{\jmath}
\safemath{\dict}{\matD}
\safemath{\inputdim}{N}		% number of columns of dictionary D
\safemath{\outputdim}{M}		%number of rows of dictionary D
\safemath{\sparsity}{S}	%sparsity
\safemath{\inputdimA}{{N_a}}	%total number of elements in dictionary A
\safemath{\inputdimB}{{N_b}}	%total number of elements in dictionary B
\safemath{\elemA}{{n_a}}	%number of elements chosen from dictionary A
\safemath{\elemB}{{n_b}}	%number of elements chosen from dictionary B
\safemath{\resA}{\matR_a}	%restriction map to elements of dictionary A
\safemath{\resB}{\matR_b}	%restriction map to elements of dictionary B
\safemath{\subD}{\matS} %subdictionary
\safemath{\subA}{\matS_a} %subdictionary part of A
\safemath{\subB}{\matS_b} %subdictionary part of B
\safemath{\dicta}{\matA} 	% first subdictionary
\safemath{\dictb}{\matB} 	% second subdictionary
\safemath{\hollowS}{H}
\safemath{\hollowA}{H_a}
\safemath{\hollowB}{H_b}
\safemath{\cross}{Z}
\safemath{\coh}{\mu_d}			% coherence dictionary
\safemath{\coha}{\mu_a}			% coherence first subdictionary
\safemath{\cohb}{\mu_b}			% coherence second subdictionary
\safemath{\mubs}{\nu}	%block sub-coherence
\safemath{\cohm}{\mu_m} %mutual coherence
\safemath{\dictset}{\setD}	% set of dictionaries
\safemath{\dictsetp}{\dictset(\coh,\coha,\cohb)}	% set of dictionaries parametrized
\safemath{\dictsetgen}{\dictset_\text{gen}}
\safemath{\dictsetgenp}{\dictsetgen(\coh)}
\safemath{\dictsetonb}{\dictset_\text{onb}}
\safemath{\dictsetonbp}{\dictsetonb(\coh)}

\safemath{\leftside}{U}
\safemath{\rightsideA}{R_a}
\safemath{\rightsideB}{R_b}

\safemath{\indexS}{\setI_S} %set of indices participating in sub-dictionary S

\safemath{\na}{n_a}			% cardinality of set of linearly independent columns of first dictionary
\safemath{\nb}{n_b}			% cardinality of set of linearly independent columns of second dictionary
\safemath{\coeffa}{p_i}	%coefficients for columns of A
\safemath{\coeffb}{q_j}	%coefficients for columns of B
\safemath{\seta}{\setP}		% set of linearly independent columns of A
\safemath{\setb}{\setQ}     % set of linearly independent columns of B
\safemath{\setw}{\setW}	%set of n largest elements of w
\safemath{\setz}{\setZ}	%set of L-n largest elements of z
\safemath{\cola}{\veca}		% generic element of the dictionary A
\safemath{\colb}{\vecb}		% generic element of the dictionary B
\safemath{\cold}{\vecd}		% generic element of the dictionary D
\safemath{\inputvec}{\vecx} 	%coefficient vector (input)
\safemath{\error}{\vece}	%error vector
\safemath{\noiseout}{\vecz} 	%noisy output vector
\safemath{\inputvecel}{x}
\safemath{\inputveca}{\vecx_a}
\safemath{\inputvecb}{\vecx_b}
\safemath{\outputvec}{\vecy}	%output of Dictionary
\safemath{\lambdamin}{\lambda_{\mathrm{min}}}
\safemath{\elltwo}{\ell_2}
\safemath{\ellone}{\ell_1}
\safemath{\ellzero}{\ell_0}
\safemath{\ellinf}{\ell_\infty}
\safemath{\ellinftilde}{\ell_{\widetilde\infty}}
\safemath{\licard}{Z(\coh,\coha,\cohb)}
\safemath{\xsol}{\hat{x}}
\safemath{\xbord}{x_b}		%Solution at the border
\safemath{\xstat}{x_s}		%Solution stationary in l0 prob
\safemath{\xstatLone}{\tilde{x}_s}
\safemath{\order}{\mathcal{O}} %order notation (big O)
\safemath{\scales}{\Theta} %scales as
\safemath{\ones}{\mathbf{1}} %all ones matrix
\safemath{\zeroes}{\mathbf{0}} %all zeroes matrix
\safemath{\thlone}{\kappa(\coh,\cohb)} %treshold l1 problem
\safemath{\constoneA}{\delta} %constant in l1 theorem to save space
\safemath{\constoneB}{\epsilon} %constant in l1 theorem to save space
\safemath{\nlarge}{L}				   %num large elements
\safemath{\sumlarge}{S_\nlarge}
\safemath{\maxlarger}{P_\nlarge}	   % maximum in Gribonval and Nielsen
\safemath{\Pzero}{\textrm{P0}}	
\safemath{\Pone}{\textrm{P1}}
\safemath{\vecfir}{\vecw}			 % \vecv element of the kernel of the dictionary, \vecv=[\vecfir \vecsec]
\safemath{\vecsec}{\vecz}
\safemath{\elvecfir}{w}              % element of vecfir
\safemath{\elvecsec}{z}				 % element of vecsec
\safemath{\nlargefir}{n}
\safemath{\normout}{\gamma}
\safemath{\auxfun}{h}
\safemath{\supp}{\textrm{supp}}%support

\safemath{\indexa}{\ell}
\safemath{\indexb}{r}
\safemath{\indexc}{i}
\safemath{\indexd}{j}

\safemath{\project}{P}%projector

\usepackage{xspace}
\newcommand{\WiFi}{\mbox{Wi-Fi}\xspace}

\usepackage{framed}

\IEEEoverridecommandlockouts
\allowdisplaybreaks % THIS ALLOWS EQUATIONS TO BREAK THE PAGE

\begin{document}

\title{A Software-Defined and Distributed \WiFi Channel-State Information Acquisition Testbed}

\author{\IEEEauthorblockN{Frederik Zumegen and Christoph Studer\thanks{This work was supported in part by an ETH Zurich Research Grant, by the Swiss National Science Foundation (SNSF) grant 200021\_207314, and by CHIST-ERA grant for the project CHASER (CHIST-ERA-22-WAI-01) through the SNSF grant 20CH21\_218704.}}\\
\IEEEauthorblockA{\em Department of Information Technology and Electrical Engineering, ETH Zurich, Switzerland\\
              email: fzumegen@iis.ee.ethz.ch and studer@ethz.ch}	
}

\maketitle

\renewcommand{\abstractname}{Abstract}
\begin{abstract}

We propose a software-defined testbed for \WiFi channel-state information (CSI) acquisition.
This testbed features distributed software-defined radios (SDRs) and a custom IEEE 802.11a software stack that enables the passive collection of CSI data from commercial off-the-shelf (COTS) devices that connect to an existing \WiFi network.
Unlike commodity \WiFi sniffers or channel sounders, our software-defined testbed enables a quick exploration of advanced CSI estimation algorithms in real-world scenarios from naturally-generated \WiFi traffic.
We explore the effectiveness of two advanced algorithms that denoise CSI estimates, and we demonstrate that CSI-based positioning of COTS \WiFi devices with a multilayer perceptron is feasible in an indoor office/lab space in which people are moving.  
\end{abstract}
%%%%%%%%%%%%%%%%%

\section{Introduction}
Localizing user equipments (UEs) in wireless systems enables location-aware services and can be utilized to improve communication.
Instead of deploying specialized positioning sensors, one can leverage the existing wireless infrastructure to perform joint communication and localization~\cite{6g_localization_sensing}.
In particular, channel-state information (CSI) estimated at the infrastructure base stations or access points (APs) describes how radio-frequency (RF) waves between each transmitting and receiving antenna interact with the environment, e.g., through reflections or scattering.
Therefore, processing CSI with geometric methods~\cite{kotaru2015spotfi, vasisht2016chronosWifiLocalization} or machine learning (ML) algorithms~\cite{Wu2018nnIndoorPos, Gönültaş2021probabilityFusion,gassner2021opencsi,wang2017indoorLocalDeepLearning,yang2013fromRSSItoCSI,arnold2019mMIMOchannelSounder} enables the extraction of location information about the transmitting UEs.
Since CSI is acquired anyway for coherent data detection, CSI-based positioning does not require the deployment of additional sensors and merely requires novel algorithms that process the already-acquired CSI.  

\subsection{CSI Acquisition Methods}

CSI acquisition can be accomplished in various ways. 
Channel sounders, such as the ones described in~\cite{dichasus2021,sandra2024widebandChSounder,bast2020cnnPositioning,arnold2019mMIMOchannelSounder}, transmit custom waveforms specifically designed to estimate CSI at high accuracy and in short time intervals with carefully calibrated hardware.
However, channel sounders do typically \emph{not} support regular bidirectional data traffic with commercial off-the-shelf (COTS) UEs and therefore, real-world communication systems will not result in CSI estimates of the same quality and volume. 
A vital step to making CSI-based positioning work in practice necessitates algorithm development and evaluation of this technology with CSI acquired from real-world COTS hardware in real-world scenarios with regular wireless traffic.  
Solutions for CSI extraction from regular wireless traffic, such as \cite{halperin2011csiIntelNIC5300,xie2015atherosCSItool}, collect CSI from commodity IEEE 802.11 \WiFi network interface cards (NICs). While easy to deploy, such tools generate CSI at rather low subcarrier resolution, with a limited number of antennas, and with fixed channel-estimation algorithms implemented on the NICs. 
To improve CSI data collection with commodity \WiFi chipsets, the recent work in~\cite{euchner2024espargos} builds phase-coherent multi-antenna \WiFi sniffers with multiple ESP32 microcontrollers \cite{esp32s2_datasheet}. While this approach enables the acquisition of CSI at multiple receive antennas and APs from regular \WiFi traffic in a cost-effective manner, one has no control over the algorithms used for CSI estimation.

\begin{figure}[t]
\centering
\includegraphics[width=\columnwidth]{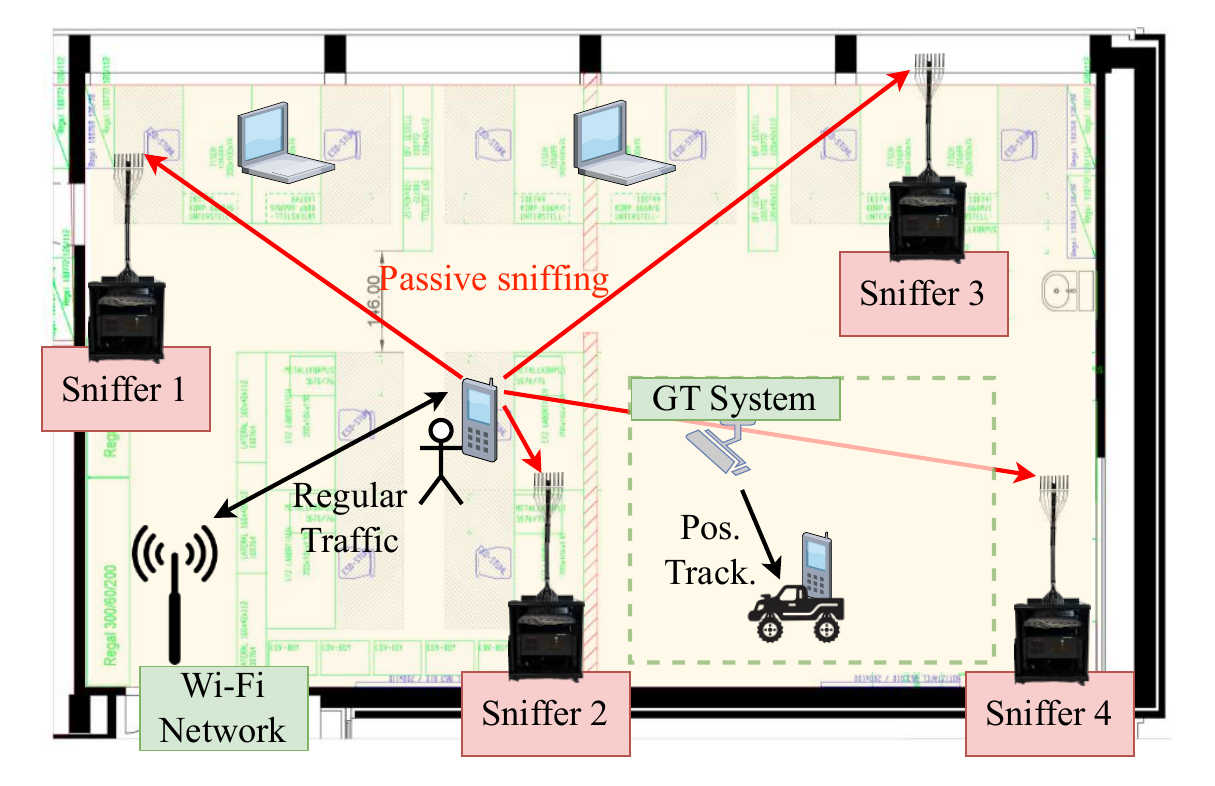}
\caption{Overview of the proposed software-defined \WiFi CSI acquisition testbed. Four distributed and software-defined \WiFi sniffers extract CSI estimates from regular IEEE 802.11a \WiFi traffic. A camera-based ground truth (GT) positioning system~\cite{worldviz} accurately tracks the transmitting objects within a confined area. This GT position information can then be used, for example, to train ML-based positioning methods.}
\label{fig:testbed-overview}
\end{figure}

\subsection{Contributions}

In contrast to channel sounders and existing \WiFi sniffers, we propose a software-defined and distributed \WiFi CSI acquisition testbed that enables multi-antenna and multi-AP CSI acquisition in realistic scenarios with real-world data traffic, while providing full control over the implemented CSI-estimation algorithms. 
Our testbed utilizes software-defined radios (SDRs) and an attached host computer with a custom baseband processing Python software stack, which enables easy exploration of CSI estimation algorithms with the goal of improving CSI-based positioning.
We describe two methods that lead to improved CSI estimates in IEEE 802.11a~\cite{802.11-2020}, namely sample-domain cyclic-prefix-aware denoising and utilizing data-carrying orthogonal-frequency division multiplexing (OFDM) symbols for channel estimation. 
We provide ML-based positioning results with CSI acquired using our testbed, demonstrating the capabilities of this software-defined testbed. 

%%%%%%%%%%%%%%%%%
\section{Software-Defined \WiFi CSI Acquisition Testbed}\label{sec:testbed}
%%%%%%%%%%%%%%%%%

\subsection{Testbed Overview}
Figure \ref{fig:testbed-overview} illustrates the concept of the proposed software-defined \WiFi CSI acquisition testbed. 
The main purpose of this testbed is to passively collect (also known as ``sniff'') CSI data from regular \WiFi traffic, i.e., from COTS \WiFi transmitters, such as phones or laptops, that connect to a regular \WiFi network. 
To this end, we place four \WiFi sniffers within the environment, each equipped with two four-antenna software-defined radios (SDRs) and a host computer. Each SDR passively records baseband in-phase/quadrature (IQ) samples at four receive antennas in the frequency band used by the transmitting COTS devices.
These baseband sample streams are then processed on the host computers with a custom software-based \WiFi receiver stack in order to extract CSI estimates.
The software stack is written entirely in Python, and we use the USRP driver \cite{ettus_uhd_manual} and the NumPy library~\cite{harris2020arrayNumpy}.
Subsequently, the locally collected CSI estimates are then combined to generate a dataset containing CSI from all four distributed \WiFi sniffer locations. This dataset can then be used to train and evaluate ML-based positioning methods.  

\subsection{\WiFi Sniffer Hardware}
\begin{figure}[tp]
    \centering
    \subfigure[]{\centering
        \includegraphics[width=0.3\columnwidth]{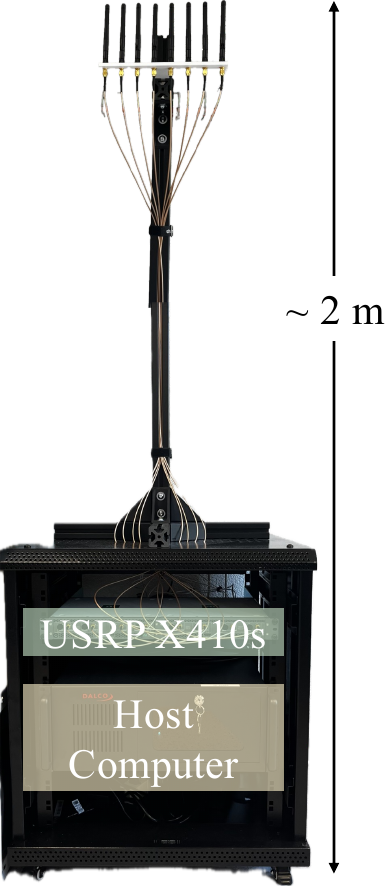}
    }    
    \subfigure[]{\centering
        \includegraphics[width=0.63\columnwidth]{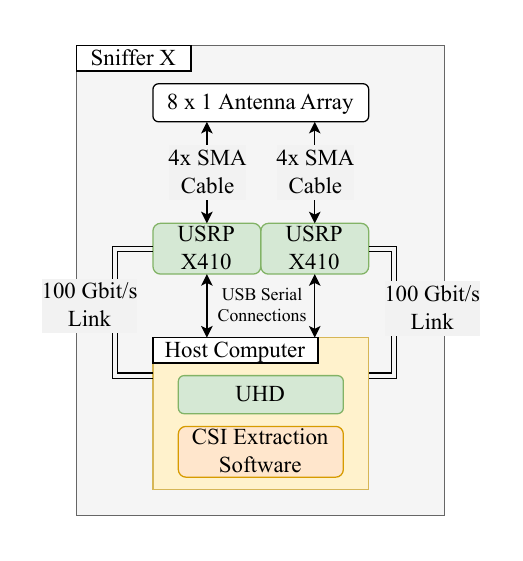}
    }
    \caption{Hardware and architecture of a \WiFi sniffer. (a) Photo of the \WiFi sniffer hardware and (b) architecture overview of our \WiFi sniffer.}
    \label{fig:sniffer-architecture}
\end{figure}

The four \WiFi sniffers are built using the hardware and architecture depicted in \fref{fig:sniffer-architecture}. 
Each \WiFi sniffer is designed to operate in the $2.4$\,GHz or $5$\,GHz ISM frequency bands---in what follows, we focus on $5$\,GHz operation. 
Eight monopole antennas of type Amphenol ST0228-30-X02-A form a uniform linear antenna array (ULA) with $2.85$\,cm antenna spacing (corresponding to approximately half the wavelength of a $5$\,GHz signal) mounted on top of each \WiFi sniffer. The antenna array height is adjustable.
Standard SubMiniature version~A (SMA) RF cables direct the wireless signals from the antenna array to two independent Universal Software Radio Peripheral (USRP) X410 SDRs~\cite{x410}.
The two USRPs convert the RF signals to baseband IQ sample streams, which are transferred to the host computer via $100$\,Gbit/s Ethernet. 
To enable fast processing of the acquired baseband samples, the host computer within each \WiFi sniffer consists of a Mellanox CX516A NIC and an Intel Core i9 CPU with 24 cores running at a maximum clock frequency of $6$\,GHz. 
For compact size and mobile deployment in an office environment, each \WiFi sniffer's housing consists of a noise-suppressed 19-inch metal rack on~wheels.

%%%
\subsection{CSI Extraction Software}

In order to extract CSI estimates from regular data traffic generated by the COTS devices that connect to an existing \WiFi network, we developed a custom software stack that processes the received baseband sample streams. 
To associate estimated CSI with the medium access control (MAC) address of the transmitting devices, we implemented an IEEE 802.11a~\cite{802.11-2020} \WiFi receiver pipeline, which not only provides CSI estimates but also attempts to decode the transmitted data within the frames.
If the cyclic redundancy check (CRC) in a \WiFi frame passes, we declare successful decoding. Furthermore, only if the transmitter's MAC address matches that of a valid\footnote{For ethical reasons, we only store CSI information from \WiFi transmitters belonging to users that gave explicit consent for data collection.} \WiFi transmitter, we store the CSI estimates along with acquired metadata locally on the sniffer's host computer.

The current implementation of the receiver pipeline only supports the decoding of frames of the non-HT format, which was introduced by the 802.11a standard. However, COTS \WiFi devices supporting newer standards, such as 802.11n or 802.11ac, still support and transmit non-HT frames.

\begin{figure*}[tp]
    \centering
    \includegraphics[width=\textwidth]{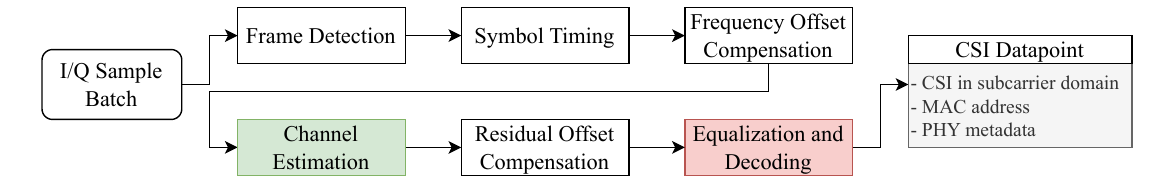}
    \caption{Physical-layer (PHY) processing pipeline executed by the CSI extraction software on the host computers. We have implemented two methods to improve CSI estimates: channel estimation (green) and equalization and decoding (red).}
    \label{fig:csi-extraction-software}
\end{figure*}

\fref{fig:csi-extraction-software} illustrates the signal flow of the custom CSI extraction software pipeline. 
The CSI extraction software's input corresponds to batches of baseband IQ samples streamed from the SDRs. A batch is a predefined number of consecutive time-domain IQ samples, e.g., $100\,000$ samples, from the four receive antennas per SDR. 
In our testbed, each USRP X410 SDR acquires four phase-synchronized baseband sample streams, which are sampled at a frequency of $40.96$\,MHz. These sample streams reach the host computer, where they are down-sampled to $20$\,MHz before being processed by our PHY software stack. In what follows, we discuss exclusively $20$\,MHz frames, which we observed to be the most common bandwidth for \WiFi frames in the non-HT format.

The \WiFi standard ensures every frame begins with a legacy preamble~\cite{802.11-2020}. The legacy preamble consists of ten short training OFDM symbols, which make up the short training field (L-STF), and two long training OFDM symbols plus a long guard interval, which make up the long training field (L-LTF).
The software stack first detects the presence and start of a \WiFi frame; this is accomplished jointly for the four phase-synchronized baseband IQ sample streams of each USRP X410 SDR using the method described by Schmidl and Cox in~\cite{schmidl1997OFDMsync}.
This frame-start detection method exploits the periodicity of the ten short OFDM training symbols in the~L-STF.

After estimating the start sample index of a \WiFi frame, we partition the acquired baseband samples into consecutive orthogonal frequency-division multiplexing (OFDM) symbols.
Then, we estimate the carrier-frequency offset (CFO) in the time domain from the two long training OFDM symbols of the L-LTF. CFO compensation is done in the time domain for each receive channel.
After initial CFO compensation and before channel estimation, the two L-LTF OFDM symbols are transformed into the subcarrier domain by a fast Fourier transform (FFT). We then perform channel estimation using these two subcarrier-domain L-LTF OFDM symbols, resulting in subcarrier domain CSI for each receive antenna; see \fref{sec:init_chest} for the details. 
For each subsequent OFDM symbol, we carry out residual phase-offset compensation (caused, e.g., by residual carrier-frequency offset, phase noise, sampling-rate offset, and Doppler shift). 

We then utilize the initial channel estimates to equalize the remaining OFDM symbols in the transmitted frame using maximum-ratio combining (MRC) with the data obtained at the four antennas per SDR. 
Next, the software estimates the \WiFi frame format of the frame currently inside the pipeline based on the modulation of the fifth, sixth, and seventh OFDM symbols; refer to~\cite{mathworksWlanFormatDetect} for the details. If this process detects a frame format other than non-HT frames, then it is discarded. Currently, our software stack only processes non-HT \WiFi frames.
For non-HT frames, the fifth OFDM symbol corresponds to the legacy signal field (L-SIG)~\cite[Chp. 17.3]{802.11-2020}, which contains the modulation and coding scheme (MCS) as well as the total length of the frame. The L-SIG field itself is encoded by a BPSK modulated rate $1/2$ convolutional code.
The remaining OFDM symbols of the frame are decoded with the MCS and frame length information from the L-SIG field. These remaining OFDM symbols are the so-called DATA field of the frame.
The decoded DATA field OFDM symbols yield the DATA bit stream. From here, we extract the MAC addresses of the transmitter and destination.

\subsection{Collected CSI Data}
The CSI estimates in the OFDM subcarrier domain along with the transmitter's MAC address and other metadata of each successfully decoded and valid frame are stored as a \textit{CSI datapoint}. The specific data per CSI datapoint is as follows:
\begin{itemize}
    \item Transmitter MAC address 
    \item Timestamp in seconds
    \item Complex-valued subcarrier-domain CSI as a 2-dimensional array: 
    \begin{itemize}
        \item Dimension~1: number of receive antennas per SDR
        \item Dimension~2: number of used OFDM subcarriers~$W_\textnormal{used}=|\Omega_\textnormal{used}|$
    \end{itemize}
    \item Estimated CFO
    \item SNR per receive antenna\footnote{The SNR per receive antenna is estimated from the  baseband IQ samples based on a pre-set power threshold that separates the received \WiFi signals from ambient noise.} 
\end{itemize}
Each \WiFi sniffer collects CSI datapoints independently. When combining the CSI estimates from all sniffers, we group these CSI datapoints into one \emph{combined CSI datapoint} whenever their timestamps are within a timing window of $30$\,ms.

In what follows, we denote the subcarrier-domain CSI of the $a$th receive antenna on subcarrier $\omega\in\Omega_\textnormal{used}$ as $\hat{h}_a[\omega]\in\complexset$. Here, $\Omega_\textnormal{used}$ denotes the set of used subcarriers (i.e., subcarriers for which we have pilot signals in the L-LTF OFDM symbols).

%%%%%%%%%%%%%%%%%
\section{CSI Denoising Methods}
\label{sec:improved_csi_esimation}
%%%%%%%%%%%%%%%%%

Large datasets that contain accurate CSI are crucial for ML-based positioning.
The key advantage of our \emph{software-defined} \WiFi CSI acquisition testbed is that we can develop and utilize advanced algorithms that lead (i) to more CSI estimates per detected \WiFi frame and (ii) to improved (i.e., more accurate) CSI than what is commonly obtained from conventional \WiFi transceivers. 
In addition, the software-defined nature of our testbed enables us to quickly evaluate the effect of different CSI-estimation algorithms on the accuracy of positioning.

In what follows, we first briefly summarize how CSI is typically extracted from the two L-LTF OFDM symbols. 
We then describe two methods (highlighted with colors in \fref{fig:csi-extraction-software}) that lead to more accurate CSI and also more CSI estimates per successfully decoded and valid \WiFi frame. 
The first method utilizes the known cyclic prefix (CP)-length of OFDM symbols to denoise subcarrier-domain CSI estimates. 
The second method exploits the successfully decoded information bits to generate one CSI estimate per OFDM symbol in the DATA field instead of only one CSI estimate using the two L-LTF symbols. 
To simplify our discussion, we omit the receive antenna index~$a$.

\subsection{Initial CSI Estimation}
\label{sec:init_chest}
CSI in the subcarrier domain is typically estimated from the two long training symbols of the L-LTF and for all used subcarriers indexed by the set $\Omega_\textnormal{used}$. Estimates of the channel's transfer function at subcarrier $\omega \in \Omega_\textnormal{used}$ are obtained as
\begin{align}\label{eq:ls-chest}
\hat{h}[\omega] = \frac{y_\textnormal{L}[\omega,1]+y_\textnormal{L}[\omega,2]}{2\>x_\textnormal{L}[\omega]}.
\end{align}
Here, $y_\textnormal{L}[\omega,1]$ and $y_\textnormal{L}[\omega,2]$ is the received signal on subcarrier~$\omega$ from the first and second L-LTF OFDM symbol, respectively, and $x_\textnormal{L}[\omega]$ is the pilot signal used on the $\omega$th subcarrier in both of these OFDM symbols.
While this initial channel estimate is typically sufficient for successful decoding of \WiFi frames, our goal is to generate more accurate CSI estimates, which have the potential to improve CSI-based positioning. 

\subsection{CP-Aware Sample-Domain CSI Denoising}
\label{sec:cp-aware-denoising}
In order to improve the accuracy of CSI estimates, we leverage the fact that \WiFi systems train more subcarriers than the length of the longest allowed impulse response---this enables denoising methods. 
Specifically, to maintain orthogonality between subcarriers, OFDM-based systems are designed for a known cyclic-prefix (CP) length $P$, which must satisfy $P\geq D_\textnormal{max}-1$ where $D_\textnormal{max}$ is the largest delay spread of the wireless channel (measured in samples).
In the L-LTF OFDM symbols, IEEE 802.11a transmits pilots on \emph{all} $W_\textnormal{used}=|\Omega_\textnormal{used}|$ used subcarriers, which is larger than the CP length $P$ (e.g., in the $20$\,MHz mode, $W_\textnormal{used}$ can be $52$ whereas $P$ can be $16$). 
This redundancy in pilots enables one to denoise estimated CSI using a procedure described in, e.g., \cite{häne2008vlsi}. 
We now detail the idea behind this approach.

One first acquires initial CSI estimates on the set of used subcarriers 
$\hat{h}[\omega]$, $\omega\in\Omega_\textnormal{used}$, e.g., using the standard procedure discussed in 
\fref{sec:init_chest}; we define the vector $\hat\bmh_\textnormal{used}\in\complexset^{W_\textnormal{used}}$ to contain only these initial CSI estimates.
It is crucial to realize that these CSI estimates are a subset of the discrete Fourier transform (DFT)-domain CSI $\hat\bmh\in\complexset^W$ given by the $W\times W$ DFT matrix $\bF$ applied to a zero-padded version of the channel's impulse response $\bmh\in\complexset^{P+1}$. Here, $W$ is the total number of subcarriers (e.g., $W=64$). 
Concretely, we have $\hat\bmh = \bF [\bmh;\bZero_{W-P-1}]$, where $\bZero_K$ is the all-zeros vector of dimension $K$ and $[\bma;\bmb]$ denotes stacking of two column vectors $\bma$ and $\bmb$. 
Since we only have initial channel estimates on the set of used subcarriers, the following relation must hold:
\begin{align}
\hat\bmh_\textnormal{used}= \bF_{\Omega_\textnormal{used},[0:P]} \bmh + \bme.
\end{align}
Here, $\bF_{\Omega_\textnormal{used},[0:P]}$ corresponds to the submatrix of the $W\times W$ DFT matrix $\bF$ with rows taken from the set of used subcarriers $\Omega_\textnormal{used}$ and the first $P+1$ columns, and $\bme\in\complexset^{W_\textnormal{used}}$ models channel-estimation noise. 
Since the number of noisy estimates in $\hat\bmh_\textnormal{used}$ is larger than the length of the sample-domain impulse response $\bmh$, we can estimate $\bmh$ from $\hat\bmh_\textnormal{used}$ using least-squares followed by computing denoised channel estimates~as follows: 
\begin{align}
\hat\bmh_\textnormal{used}^\textnormal{DN}  = \bF_{[0:W-1],[0:P]} \bF^\dagger_{\Omega_\textnormal{used},[0:P]} \hat\bmh_\textnormal{used}.
\end{align}
Here, $\bF_{[0:W-1],[0:P]}$ is the submatrix of the $W\times W$ DFT matrix~$\bF$ with the first $P+1$ columns and the superscript~$^\dagger$ denotes the Moore-Penrose pseudo-inverse \cite{laub_mppseudoinverse}. 
Note that $\hat\bmh_\textnormal{used}^\textnormal{DN}\in\complexset^{W}$ contains CSI estimates on all $W$ subcarriers, even those for which no pilots were used. Furthermore, $\hat\bmh_\textnormal{used}^\textnormal{DN}$ is a denoised version of the initially-estimated CSI.

\subsection{Channel Estimation on All OFDM Symbols}\label{sec:all_symbol_chest}
For non-HT \WiFi frames as specified in IEEE 802.11a, only two OFDM symbols in the L-LTF are used for channel estimation; see~\fref{sec:init_chest}.
We now propose a method that yields channel estimates for \emph{every} transmitted OFDM symbol beyond the preamble (L-SIG field and DATA field), resulting in more CSI estimates. These additional CSI estimates can, for example, be used to further denoise the subcarrier CSI.\footnote{Other uses might be to extract Doppler shift information over the entire duration of a \WiFi frame, which we leave for future work.}
We now detail the idea behind this approach.

When a frame is successfully decoded, i.e., if the CRC of the DATA field indicates successful decoding, we assume that the decoded bit stream corresponds to the true transmitted bits from the DATA field.
One can now reverse the PHY processing pipeline and transform these transmitted data bits back to constellation symbols, e.g., 16-QAM, BPSK, etc., on the corresponding subcarriers and within the appropriate OFDM symbols.
These known constellation symbols can then be used as pilot signals in order to estimate the channel's transfer function on every used subcarrier for every OFDM symbol in the L-SIG \emph{and} DATA fields. 

Let $s[\omega,\ell]$ be the constellation symbol on the $\omega$th subcarrier for the $\ell$th OFDM symbol obtained by reversing the PHY pipeline as outlined above; assume that $s[\omega,\ell]$ is the true transmitted constellation symbol. Then, the input-output relation on this subcarrier and for this OFDM symbol is given by
\begin{align}
y[\omega,\ell] = \hat{h}[\omega] s[\omega,\ell] + n[\omega,\ell].
\end{align}
One can now extract an estimate of the channel's transfer function $\hat{h}[\omega]$ as
\begin{align} \label{eq:persymbolestimate}
\hat{h}[\omega,\ell] = \frac{y[\omega,\ell]}{s[\omega,\ell]},
\end{align}
which is the channel estimate for the $\omega$th subcarrier and the $\ell$th OFDM symbol in the L-SIG plus DATA field. 

In what follows, we are interested in combining all of the channel estimates $\hat{h}[\omega,\ell]$, $\ell=1,\ldots,L$, with $L$ being the number of OFDM symbols in the L-SIG plus DATA field, in order to further denoise the subcarrier CSI. 
Since the estimates in \fref{eq:persymbolestimate} are obtained through different constellation symbols with potentially different magnitude (e.g., for 16-QAM, not all constellation points have the same magnitude), one should \emph{not} simply average these estimates for denoising. 
In fact, estimates obtained from constellation symbols with larger magnitude result in smaller estimation error. 
Mean-square error (MSE)-optimal combining can be obtained by following the approach put forward in \cite[Sec. IV-A]{jeon2019decentralized}, which results in the following channel estimate for each used subcarrier $\omega\in\Omega_\textnormal{used}$:
\begin{align} \label{eq:averageddenoising}
\hat{h}[\omega] = \frac{1}{\sum_{\ell'=1}^L  |s[\omega,\ell']|^2} \sum_{\ell=1}^L  |s[\omega,\ell]|^2 \hat{h}[\omega,\ell].
\end{align}
This approach puts more importance to estimates in \fref{eq:persymbolestimate} obtained through constellation symbols with more energy (e.g., the four corner symbols in 16-QAM), whereas estimates obtained through weaker symbols are de-weighted. 
Note that one can easily incorporate the two estimates generated during the L-LTF frame to further denoise the estimated CSI $\hat{h}[\omega]$. 

%%%%%%%%%%%%%%%%%
\section{Results}\label{sec:results}
%%%%%%%%%%%%%%%%%\

\begin{figure}[tp]
    \centering
    \includegraphics[width=0.9\linewidth]{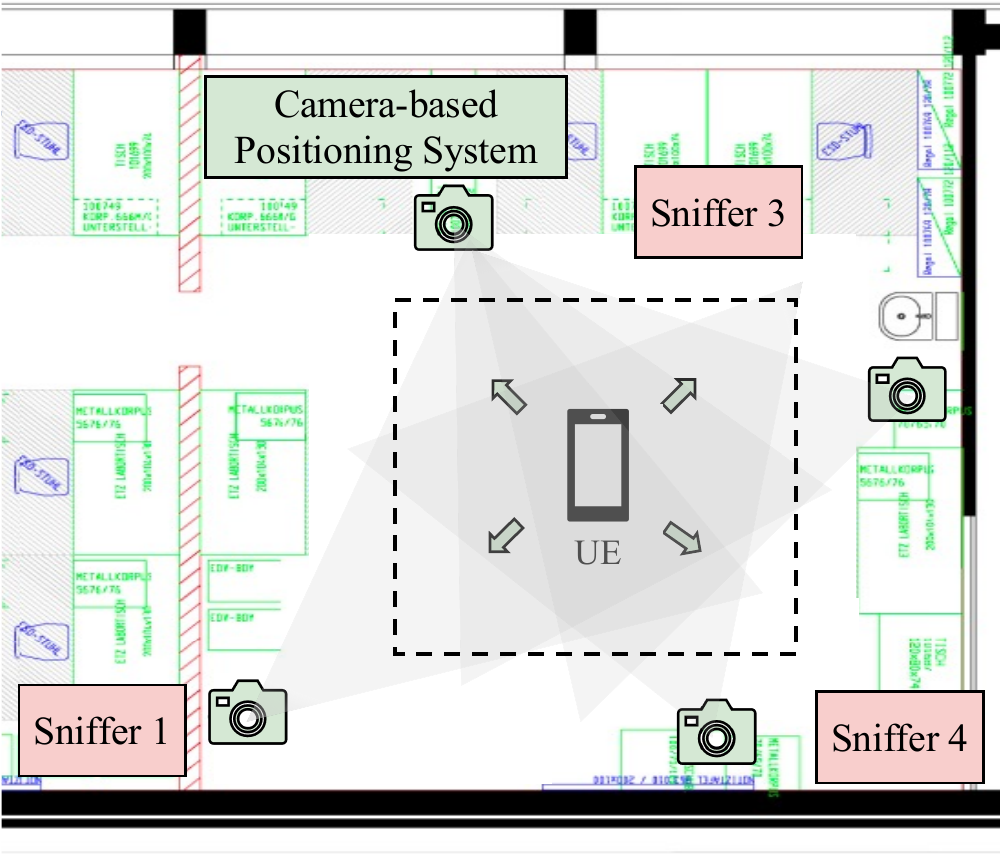}
    \caption{Floorplan of the measurement area (black dashed lines) in the shared office/lab space. A robot platform moves the UE around the measurement area in randomized fashion. The GT positioning system collects accurate position information alongside the CSI data.}
    \label{fig:measurement-scenario}
\end{figure}

\begin{figure*}[tp]
    \centering
    \subfigure[]{
        \includegraphics[width=0.32\textwidth]{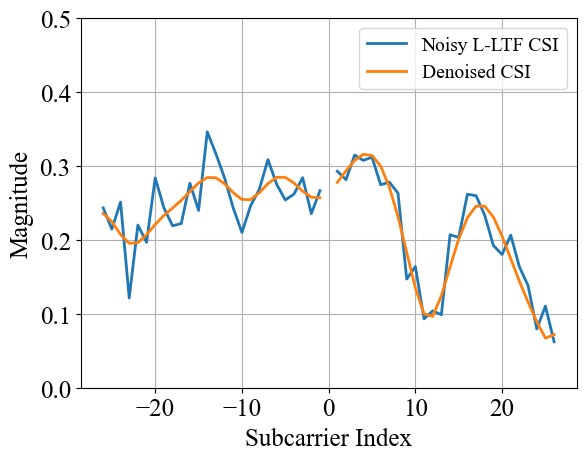}
        \label{fig:single-frame-csi-comparison}
    }
    \hfill
    \subfigure[]{
        \includegraphics[width=0.28\textwidth]{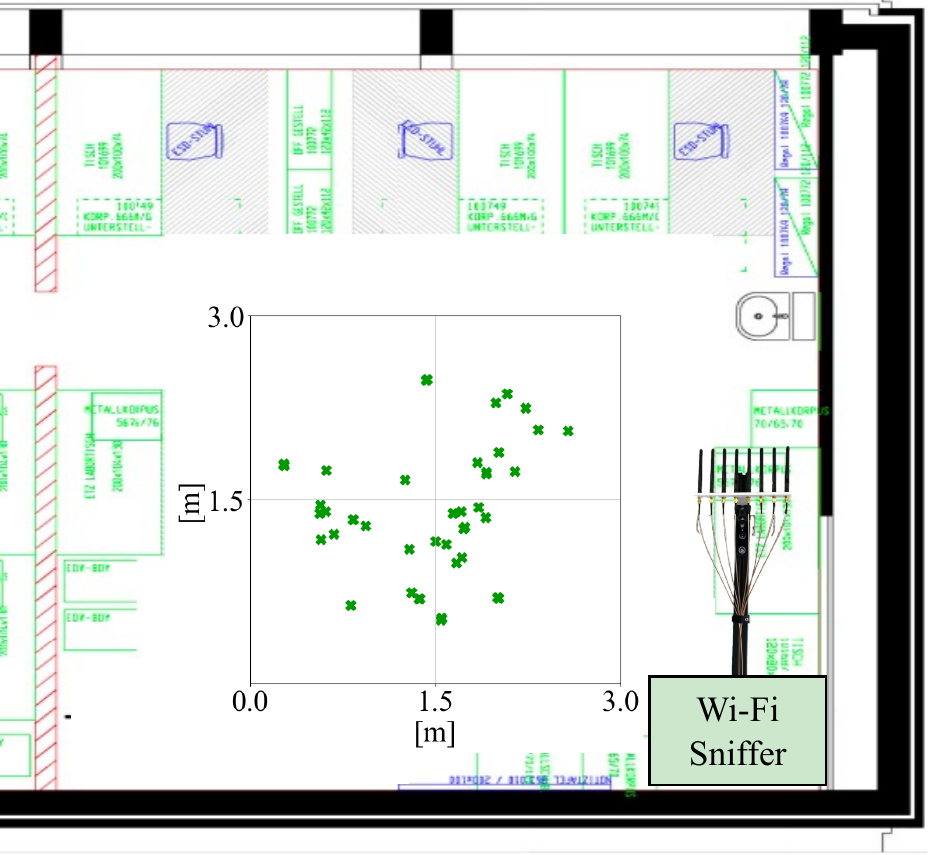}
        \label{fig:measured-static-tx-locations}
    }
    \hfill
    \subfigure[]{
        \includegraphics[width=0.32\textwidth]{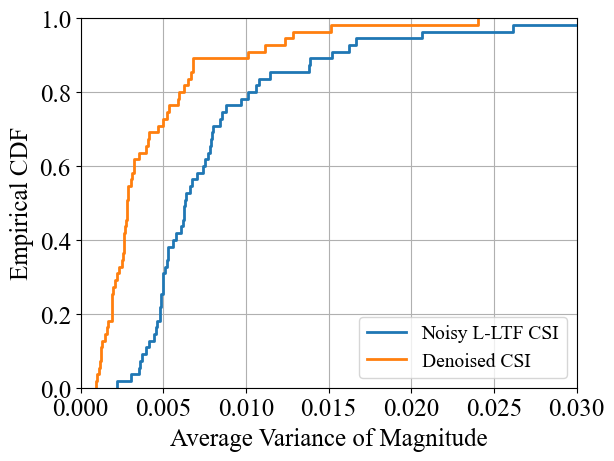}
        \label{fig:csi-variance}
    }
    \caption{(a) A single frame's CSI is improved by combining the denoising methods described in \fref{sec:cp-aware-denoising} and \fref{sec:all_symbol_chest}. (b) At these fixed transmit locations (green) at least $20$ CSI datapoints were measured to compare the variance between initial CSI estimates and denoised CSI. (c) For fixed transmitter locations, the variance of the magnitude is considerably lower (better) for denoised CSI than for the CSI estimates obtained from the two L-LTF OFDM symbols only.}
    \label{fig:csi-variance-results}
\end{figure*}

We now evaluate the effectiveness of the proposed CSI-estimation methods described in \fref{sec:improved_csi_esimation}. We then show end-to-end results for CSI-based positioning using ML with CSI data obtained from our testbed. 
The CSI data collected used in the subsequent evaluation was acquired in an area of approximately $4\,\textnormal{m} \times 4\,\textnormal{m}$ inside a shared office/lab space (see \fref{fig:measurement-scenario}). 
During a CSI acquisition campaign, a single UE generates data traffic forced by the user datagram protocol (UDP), such that it would regularly transmit (and receive) \WiFi frames. The UE used in the sample acquisition campaigns was an Apple iPhone 14 Pro.
The UE was mounted onto a robot platform and guided through the measurement area in a randomized fashion; the UE height was fixed to $0.2$\,m. Note that people were working (and moving) in the shared office/lab space during CSI acquisition.

\begin{figure*}[tp]
    \centering
    \subfigure[]{
        \includegraphics[width=0.31\textwidth]{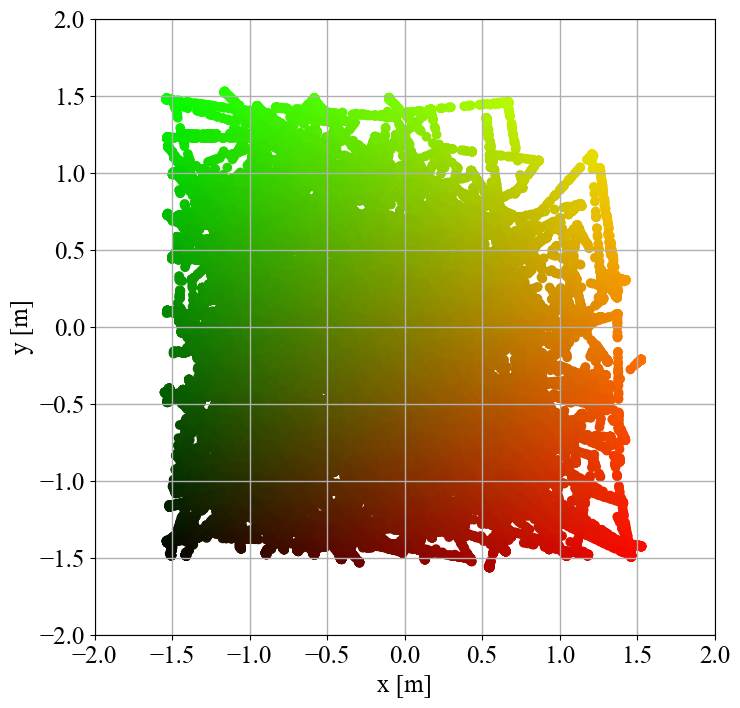}
        \label{fig:better-csi-testset}
    }
    \hfill
    \subfigure[]{
        \includegraphics[width=0.31\textwidth]{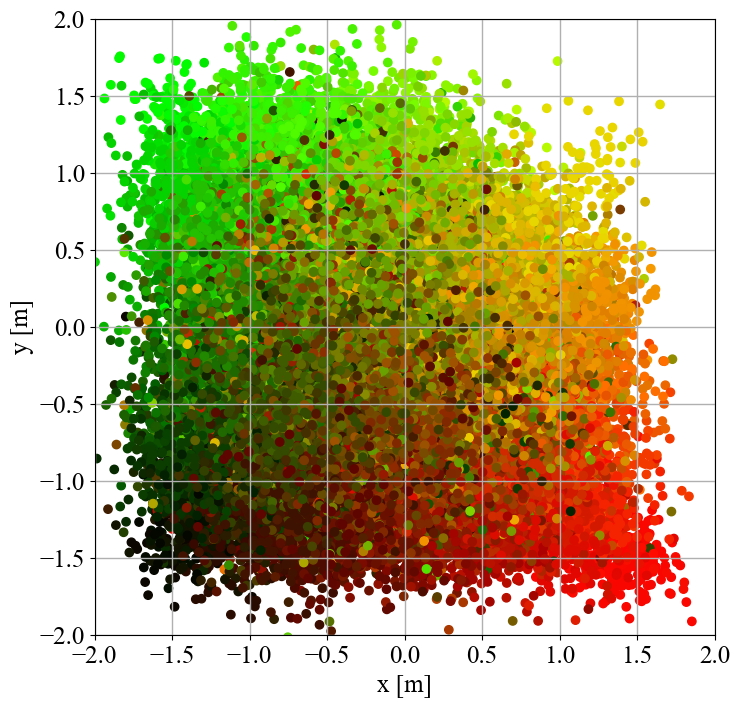}
        \label{fig:H_leg-output}
    }
    \hfill
    \subfigure[]{
        \includegraphics[width=0.31\textwidth]{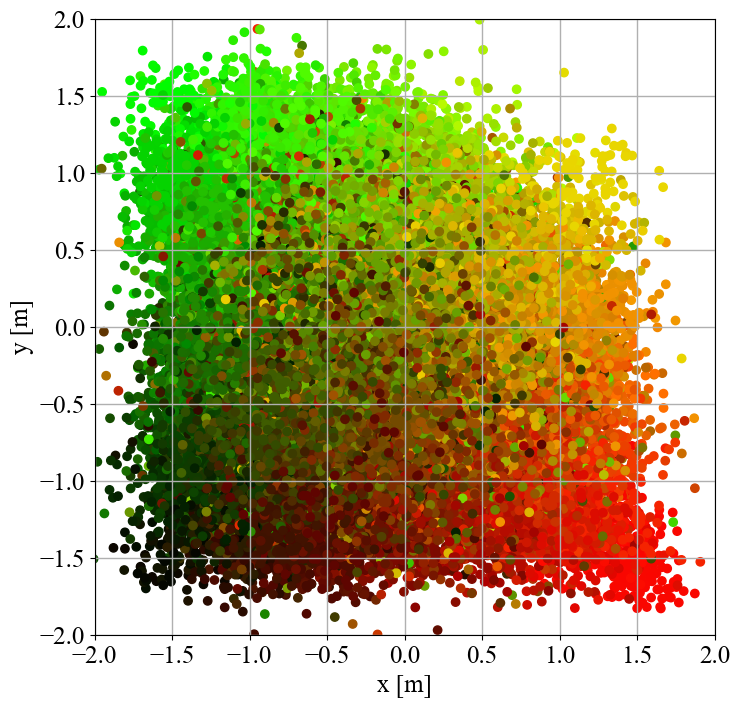}
        \label{fig:H_data-output}
    }
    \caption{(a) Ground-truth positions of the test set; (b)  evaluation of ML-based positioning on the test set with CSI from L-LTF symbols only; (c) evaluation of ML-based positioning on the test set with improved CSI from CP-aware denoising and L-LTF, L-SIG, and DATA field weighted-average CSI.}
    \label{fig:positioning-better-csi}
\end{figure*}

\subsection{Improved CSI Estimates}
Estimating CSI by performing CP-aware sample-domain denoising (as described in \fref{sec:cp-aware-denoising}) and denoising using all OFDM symbols (i.e., from the two L-LTF symbols and all symbols in the L-SIG and DATA fields as described in \fref{sec:all_symbol_chest}) results in CSI estimates that are significantly less noisy.
\fref{fig:single-frame-csi-comparison} shows a CSI estimate over the used subcarriers from the L-LTF only (blue curve) and using the proposed denoising approaches (orange curve). Evidently, combining CP-aware denoising with averaged estimates from the two L-LTF, the L-SIG, and the DATA symbols as in \fref{eq:averageddenoising} results in visibly less noisy CSI.

To further investigate the effectiveness of our CSI denoising techniques, we acquired CSI estimates from $55$ different fixed transmitter locations, indicated as green dots in \fref{fig:measured-static-tx-locations}.
When keeping the transmitter at a fixed location, we expect only a small variance in the measured subcarrier CSI (except for effects caused by moving people in the shared office/lab space).
With the proposed CSI denoising methods, the variance over multiple CSI datapoints from a fixed transmit location decreases slightly compared to the initial CSI estimation method; this becomes particularly apparent when observing the variance of the magnitude of the subcarrier CSI (see \fref{fig:csi-variance}).

\subsection{End-to-End Positioning Performance}
To demonstrate the feasibility of ML-based positioning with CSI acquired using our testbed, we trained a simple ML model with a sample dataset. The dataset's parameters are summarized in \fref{tbl:sample-dataset-params}.
In the measurement campaign for the sample dataset, we operated four \WiFi sniffers at distributed locations and with only one active SDR.
The sample dataset consists of two sets of CSI: the first set contains CSI from only the initial channel estimates (refer to \fref{sec:init_chest}); the second set contains denoised CSI (refer to \fref{sec:cp-aware-denoising} and \fref{sec:all_symbol_chest}).
This means we have two datasets for training an ML-based positioning pipeline: one for the initial CSI and one for the denoised CSI.
The ML model for the positioning experiment consists of a single, fully-connected multilayer perception (MLP).
Each combined CSI datapoint is preprocessed into a CSI feature vector before passing it to the MLP. We take the absolute values of the subcarrier CSI and scale the resulting feature vector to unit Euclidean norm. 
The neural network's parameters are summarized in \fref{tbl:positioning-nn-params} and the training strategy in \fref{tbl:positioning-training-params}. 

\fref{fig:positioning-better-csi} shows the ground-truth positions as well as two positioning results on the test set with the two sub-datasets and two different positioning MLPs.
\fref{tbl:better-csi-positioning-stats} shows the mean positioning error (in mean Euclidean distance) and the $95$\% positioning error. As a baseline, we also include the positioning error achieved by minimizing the mean positioning error of a single position estimate to all measured positions in the dataset.
As it can be seen in \fref{tbl:better-csi-positioning-stats}, ML-based positioning for both sub-datasets achieves a mean positioning error of slightly more than $50$\,cm, which is substantially more accurate than the baseline, but with many outliers visible---the estimated positions on the test set shown in \fref{fig:H_leg-output} and \fref{fig:H_data-output} are also similar. 
Quite surprisingly, the denoised dataset does not consistently improve the positioning error. We address this issue to the fact that we do not have control over the antenna orientation, which is currently dominating the positioning error and causes a large number of outliers. 
We expect that more sophisticated neural network architectures and self-supervised training techniques~\cite{studer2018cc} will be necessary to fully leverage more accurate CSI.

\begin{table}[tp]
    \caption{Dataset parameters for ML-based positioning}
    \label{tbl:sample-dataset-params}
    \centering
    \begin{tabular}{@{}ll@{}}
        \toprule
        \WiFi standard       & IEEE 802.11a            \\
        \midrule
        Bandwidth      & $20$\,MHz,     
                             $52$ used subcarriers      \\
        \midrule
        Rx hardware         & $4$ distributed sniffers, 1 SDR per sniffer, \\
                            & $4$ antennas per SDR \\
        \midrule
        Training CSI data   & $296\,892$ datapoints  \\
        \midrule
        Testing CSI data    & $74\,222$ datapoints   \\
        \bottomrule
    \end{tabular}
\end{table}

\begin{table}[tp]
    \caption{MLP parameters}
    \label{tbl:positioning-nn-params}
    \centering
    \begin{tabular}{@{}ll@{}}
        \toprule
        Number of layers       & Input, 3 hidden, output            \\
        \midrule
        Layer dimensions      & $832$, $832$, $512$, $256$, $64$, and $2$  \\
        \midrule
        Activations           & ReLU (except output layer, which is linear) \\
        \bottomrule
    \end{tabular}
\end{table}

\begin{table}[tp]
    \caption{MPL training parameters}
    \label{tbl:positioning-training-params}
    \centering
    \begin{tabular}{@{}ll@{}}
        \toprule
                Loss function       & Mean squared error           \\
                \midrule
        Number of epochs       & $50$            \\
        \midrule
        Learning rate      & $10^{-4}$     \\
        \midrule
        Optimizer      & Adam, with step-size decay \\
        \midrule
        Decay period & $20$ epochs \\
        \bottomrule
    \end{tabular}
\end{table}

\begin{table}[tp]
    \caption{End-to-end positioning performance}
    \label{tbl:better-csi-positioning-stats}
    \centering
    \begin{tabular}{@{}lcc@{}}
        \toprule
        & Mean pos.~error & $95$\% pos.~error \\
        \midrule
        Denoised CSI  & $0.53$\,m & $1.44$\,m \\
        L-LTF CSI only         & $0.54$\,m & $1.43$\,m \\
        Single best guess      & $0.94$\,m & $1.64$\,m \\
        \bottomrule
    \end{tabular}
\end{table}

%%%%%%%%%%%%%%%%%
\section{Conclusions, Limitations, and Outlook}\label{sec:conclusions}
%%%%%%%%%%%%%%%%%

We have proposed a software-defined \WiFi CSI acquisition testbed that enables multi-antenna and multi-AP CSI acquisition in realistic scenarios with real-world wireless traffic.
The software-defined nature of this testbed enables a quick exploration of advanced CSI estimation algorithms with the aim of improving CSI-based positioning techniques. 
We have explored two denoising methods that improve the quality of CSI estimates, and we have demonstrated that ML-based positioning is possible with CSI data acquired from COTS \WiFi devices that connect to an existing network.

The current testbed implementation suffers from two main limitations. First, we only support the non-HT format of IEEE 802.11a \WiFi with a bandwidth of $20$\,MHz. Supporting larger bandwidths would (i) enable the acquisition of more data points and (ii) result in higher frequency resolution, which is likely to improve positioning accuracy further. Supporting these modes is part of ongoing work. 
Second, since our testbed sniffs regular \WiFi traffic originating in the surrounding area, we have no control over the transmitting device, i.e., neither the type nor its orientation, transmit power, or beamforming strategy, and no control over the movement in the shared office/lab space. We have observed that different devices result in different CSI and even minor rotations to the transmitting device can result in vastly different CSI. 
We believe that self-supervised channel-charting-based positioning techniques~\cite{studer2018cc,taner2023ccRealWorld} have the potential to overcome these challenges. 

\balance

\bibliographystyle{IEEEtran}
\bibliography{bib/VIPabbrv,bib/confs-jrnls,bib/publishers,bib/my_bib}

\balance

\end{document}